# Modelling and analysis of biological joints of a human arm in different movements

Amit Kumar Bedaka, Ponnusamy Pandithevan

Department of Mechanical Engineering, Indian Institute of Information Technology Design and Manufacturing Kancheepuram, Chennai - 600 127, Tamilnadu, India, Email: ppthevan@iiitdm.ac.in

**Abstract-** In the present work, irregular complex joints of the human arm are modelled in the computer aided design environment. Accurately the geometric models of the joints and surrounding bony regions are developed from computed tomography data set and the range of movements are analysed. In this present work, biological joints are approximated into the mechanical joint without losing any kinematic behaviour and degrees of freedom. Then the analysis is done to determine the importance of joint geometries with different loading and boundary conditions. The present study helps to predict the behavior of joint configuration in the design stage.

1. INTRODUCTION

Human arm is a structure that can perform numerous operations without any complexity. Researchers across the globe are studying the best possible methods to mimic it. The full resemblance of the human arm development is still a daunting challenge. Human arm properties, load carrying capacity and stresses is too studied for various purposes. It can be applied to the design of the prosthesis and surgical simulation.

Mechanics related to biology, uses the principles of mechanics for solving problems associated with the structure and function of living organisms. It is the science that deals with forces and their effects, applied to biological systems. In order to design and develop the biomimetic human arm, the study of the mechanics of the human arm is important. Advantages and limitations of primary imaging modalities that make use of in different applications include, computed tomography (CT), magnetic resonance imaging (MRI), optical microscopy, micro CT, etc. described [1]. The kinematics and dynamics of the human arm and to provide the engineering specification to facilitate the design of a seven degree of freedom powered exoskeleton. The results indicate that the various joint kinematics and dynamics change significantly based on the nature of the task [2]. Flexion/extension and pronation/supination moment arms of the brachioradialis, biceps, brachialis, pronator teres, and triceps were calculated from measurements of tendon displacement and joint angle in two anatomic specimens and were estimated using a computer model of the elbow joint [3]. A computer modelling technique for predicting passive elbow flexion, extension range of motion based on impingement of bony geometry [5]. Computational model of the elbow and forearm behavior was dictated to develop and validate [6]. The kinematics and morphological data used to produce the 3D animated models from the experimental data [7]. Mechanical behavior of 42 fresh human cadaver lumbar motion segments in flexion, extension, lateral bending and torsion is examined for different loading conditions [4]. In the present work, irregular complex three dimensional human arm is modeled using the CAD software's, the biological joint were approximated into the mechanical joint without losing any kinematic behaviour and DOF.

2. METHODOLOGY

A CAD model of the human arm with the shoulder and elbow joint was constructed by acquiring computed tomography (CT) from medical images. In the present work, different methods are evaluated and compared to generating an accurate CAD model from the medical CT data set. In general, image based CAD modelling methodology involves three major steps: (1) Scanned file of human bones in form of curves; (2) Import the points file into the CAD software; and (3) Join the points and model into solid part.

The Shoulder joint has three degrees of freedom and it allows six types of rotation movements, which are abduction, adduction, flexion, extension, internal rotation and external rotation. The elbow has one degree of freedom and it allows two types of movement, which are flexion and extension. The Elbow joint is also associated with pronation and supination movement. Bones are modelled with soft tissue and bone marrow to mimic biologically. The joint of the human arm are analyses for the different movements of joints for various ranges of motion with different loading and boundary conditions [3], [11], [12].

*Sub-assembly of the human arm*

The human arm contains different types of bones each bone unique in its geometry as shown in Fig. 1. The complete individually modeled parts are imported in the Autodesk Inventor® and both bones assembled by elbow joint as like biological joint. The complete length of the human arm is 640mm from the shoulder joint to the end of the ulna bone. After complete assembly as shown in Fig. 2, the model is exported in Initial Graphics Exchange Specification (IGES) format for further analysis.

*Main Assembly of Soft Tissue, Bone and Bone Marrow*

The main assembly of the human arm consists of soft tissue, bone and bone marrow. The bones are imported individually into the assembly to model the soft tissue and bone marrow using Boolean operations as shown in Fig. 3, as like human arm.

Cross sectional view as shown in Fig .4 has the three different parts of human arm, where outer part is soft tissue followed by the bone and the inner part is bone marrow. In general, the human arm consists of soft tissue, bone and bone marrow. The modelling of the human arm is not like regular geometries. The human bone is modeled from point cloud using the reverse engineering method by adapting the structure of a real bone. The joints of the human arm are modeled individually to achieve the desired degrees of freedom as like the human arm. The soft tissue and bone marrow is modeled approximately by keeping bone as the reference.

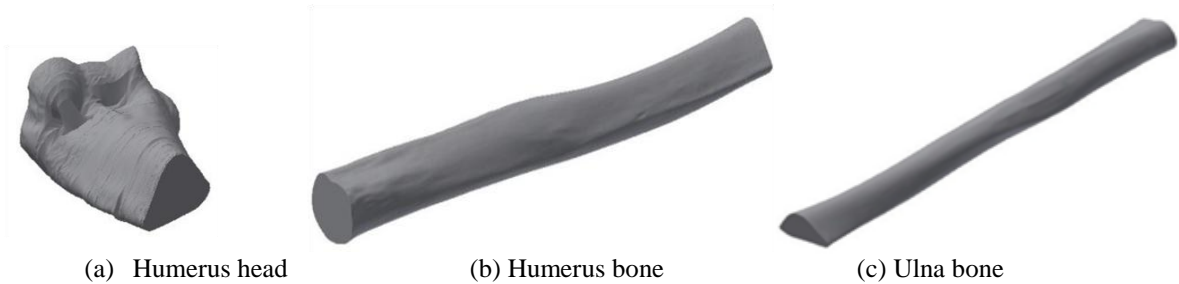

(a)  Humerus head        (b) Humerus bone        (c) Ulna bone

Figure 1. Parts of human arm.

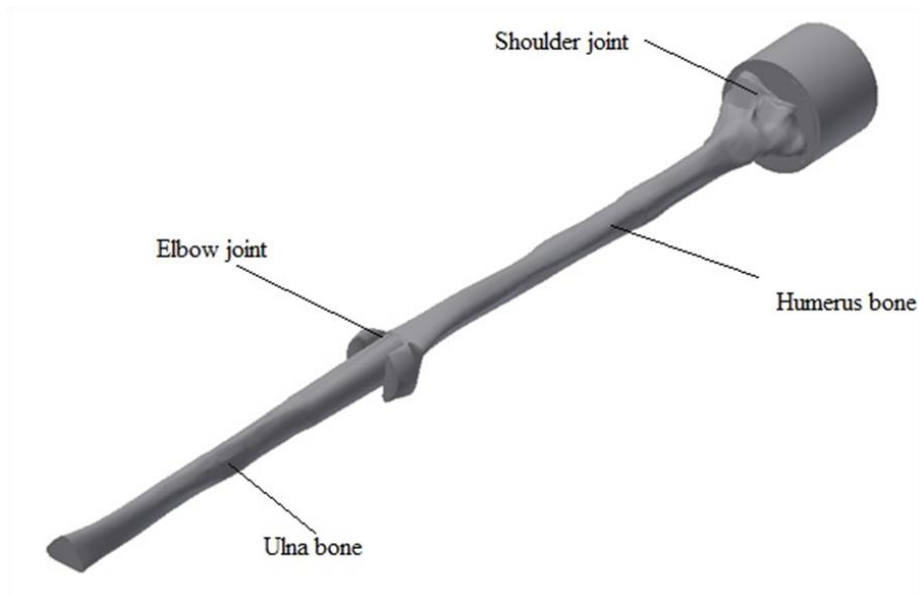

Figure 2. Sub- assembly of human arm.

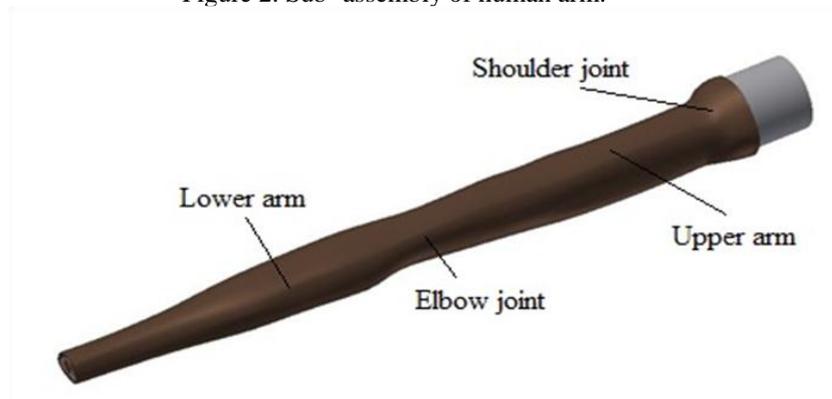

Figure 3. Complete assembly of the soft tissue, bone and bone marrow.

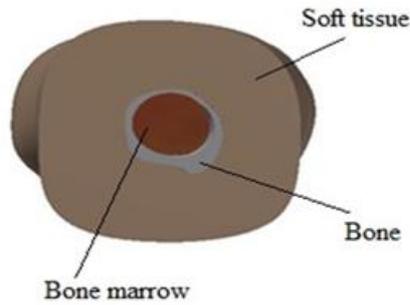

Figure 4. Cross sectional view of the complete assembly

3. RESULTS AND DISCUSSION

Range of motion was computed accurately by the model, with joint motion constrained by bone marrow, bone and software tissue. The study of biological joint is performed to understand the behaviour for different loading and boundary condition [3], [11], [12]. Material properties are mapped individually biologically. The shoulder joint and elbow joint perform the various movements under consider loading. This analysis provides useful information to predict the possible failure locations in the human arm.

Analysis of the shoulder and elbow joint with different movements for various ranges of motion is performed to measure different stresses and strains using ANSYS. From the results, it is sure that the different movements are influenced by the joints. Also, it help to understand the location of joints to get accurate range of motion. In the current study only pronation movement results are shown in Fig. 6 to understand and study the importance of biological joints in human arm. The humerus and ulna bone is assembled by elbow joint to perform the pronation movement with the given boundary condition, as shown in Table. I. The loading and boundary conditions are given with respect to the coordinate axis shown the Fig. .The shoulder is fixed and elbow joint is rotated along the z-axis. The moment 7300Nmm is applied along the z-axis at the tip of ulna bone as shown in Fig 5. Before this convergence study of the complete assembly is performed , meshed with tetrahedral mesh type of mesh size 1.25mm to perform the analysis. For given boundary and loading condition static analysis of complete assembly is performed to study the stress and strain, as shown in Table. II. The material properties of bone marrow, bone and soft tissue is assigned according to the values found by [8], [9], [10].

Table 1. Boundary Condition of Pronation Movement of Elbow Joint with fixed shoulder Joint

| Serial No. | Co-ordinate | Degree |
|---|---|---|
| 1 | Rx | $0^0$ |
| 2 | Ry | $0^0$ |
| 3 | Rz | $80^0$ |

Table 2. Stress and Strain for Pronation Movement of Elbow Joint with fixed shoulder Joint

| Serial No. | Stress/Strain | MPa |
|---|---|---|
| 1 | Equivalent Stress | 162.8 |
| 2 | Equivalent Elastic Strain | 8.388 |

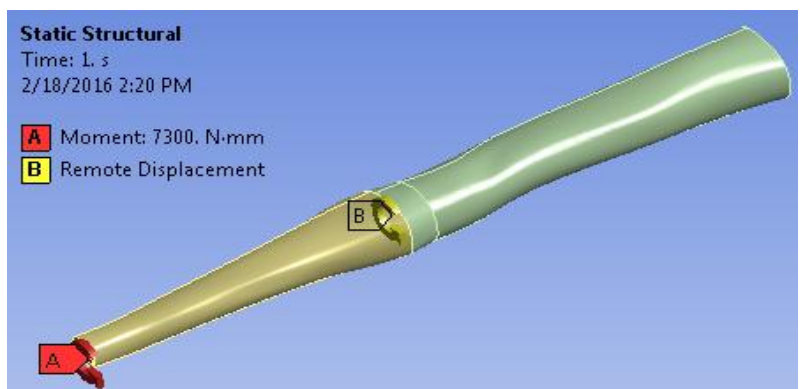

Figure 5. Boundary conditions for pronation movement

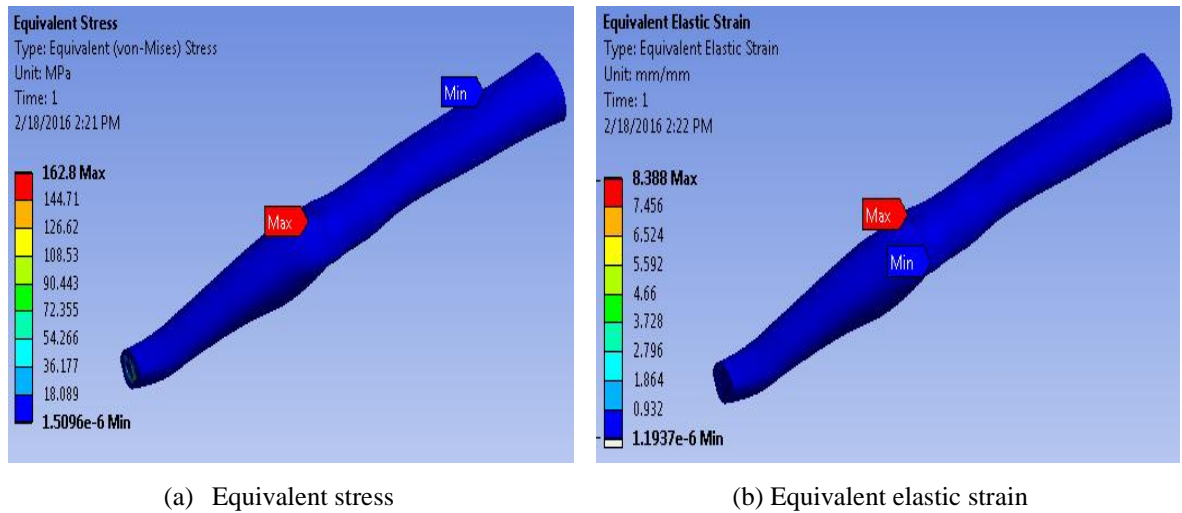

(a) Equivalent stress       (b) Equivalent elastic strain

Figure 6. Static analysis of the complete assembly

## 4. CONCLUSION

The current work presents a method to model an irregular complex joint of the human arm. The surrounding bony region was developed accurately to analyse different range of motions. In this work, biological joints are approximated into the mechanical joint without losing any kinematic behaviour and degrees of freedom. Abduction, adduction, flexion, extension, internal rotation and external rotation of the shoulder joint, whereas flexion, extension, pronation and supination from elbow joint are different movements studied for various loading and boundary conditions. The failure region was identified by the various movements of human arm associated with the shoulder and elbow joint. This analysis was done to determine the importance of joint geometries to predict the behavior of joint configuration in the design stage. These CAD model can be used to study the anatomy to make the prosthesis, also it can help in the surgery for planning.

## 5. REFERENCES

[1] W. Sun, A. Darling, B. Starly, J. Nam, "Computer-aided tissue engineering: overview, scope and challenges", Biotechnology and applied biochemistry, 39(1), 29–47, 2004.
[2] B. Hannaford, J. C. Perry, N. Manning, J. Rosen and S. Burns, "The human arm kinematics and dynamics during daily activities-toward a 7 DOF upper limb powered exoskeleton", ICAR'05.IEEE, Proceedings 12th International Conference on Advanced Robotics, 532-539, 2005.
[3] W. M. Murray, S. L. Delp, T. S. Buchanan, and M . Wendy, "Variation of muscle moment arms with elbow and forearm position." Journal of biomechanics 28(5), 513-525, 1995.
[4] A. B. Schultz, D. N. Warwick, M. H. Berkson and A. L. Nachemson, "Mechanical properties of human lumbar spine motion segments Part I: Responses in flexion, extension, lateral bending, and torsion", Journal of Biomechanical Engineering, 101(1), 46-52, 1979.
[5] R. T. Willing, M. Nishiwaki, J. A. Johnson, G. J. King and G. S. Athwal, "Evaluation of a computational model to predict elbow range of motion ", Computer Aided Surgery, 19(4), 56-63, 2014.
[6] J. P. Fisk, J. S. Wayne, "Development and validation of a computational musculoskeletal model of the elbow and forearm", Annals of biomedical engineering, 37(4), 803-812, 2009.
[7] S. V. Jan, P. Salvia, I. Hilal, V. Sholukha, M. Rooze and G. Clapworthy, "Registration of 6-DOFs electrogoniometry and CT medical imaging for 3D joint modelling", 35(11), 1475-1484, 2002.
[8] A. A. Zadpoor, "Finite element method analysis of human hand arm vibrations", Int. J. Sci. Res, 16, 391-395, 2006.
[9] A. B. Mathura, A. M. Collinswortha, W. M. Reicherta, W. E. Krausb, G. A. Truskeya, "Endothelial, cardiac muscle and skeletal muscle exhibit different viscous and elastic properties as determined by atomic force microscopy", Journal of biomechanics, 34(12), 1545-1553, 2001.
[10] W. M. Vannah and D. S. Childress, "Modelling the mechanics of narrowly contained soft tissues: the effects of specification of Poisson's ratio", Journal of rehabilitation Research and development, 30, 205-205, 1993.
[11] W. Abend, E. Bizzi and P. Morasso, "Human arm trajectory formation", Brain: a journal of neurology, 105(Pt 2), 331-348, 1982.
[12] J. L. Askew, A. N. Kainan, B. F. Morrey, E. Y. S. Chao, "Isometric elbow strength in normal individuals", Clinical orthopaedics and related research, 222, 261-266, 1987.